\documentclass[a4paper]{VisionStyle}
\usepackage{epsfig}

\newcommand {\ie} {{\it  i.e.}}

\newcommand {\eg} {{\it e.g.}}
\newcommand {\etal} {{\it et~al.}}

\newcommand {\go} {\mathrel{\hbox{\rlap{\lower.55ex \hbox {$\sim$}}
         \kern-.3em \raise.4ex \hbox{$>$}}}}
\newcommand {\lo} {\mathrel{\hbox{\rlap{\lower.55ex \hbox {$\sim$}}
         \kern-.3em \raise.4ex \hbox{$<$}}}}
\def\Msun{\ifmmode M_\odot \else $M_\odot$\fi} 
\def\Mdot{\ifmmode \dot M \else $\dot M$\fi} 
\def\hmxb{\ifmmode n_{\rm HMXB} \else $n_{\rm HMXB}$\fi}
\def\psnb{\ifmmode n_{\rm PSNB} \else $n_{\rm PSNB}$\fi}
\def\lmxb{\ifmmode n_{\rm LMXB} \else $n_{\rm LMXB}$\fi}
\def\mrp{\ifmmode n_{\rm MRP} \else $n_{\rm MRP}$\fi}
\def\tauh{\ifmmode \tau_{\rm HMXB} \else $\tau_{\rm HMXB}$\fi}
\def\taup{\ifmmode \tau_{\rm PSNB} \else $\tau_{\rm PSNB}$\fi}
\def\tauel{\ifmmode \tau_{\rm LMXB} \else $\tau_{\rm LMXB}$\fi}
\def\taum{\ifmmode \tau_{\rm MRP} \else $\tau_{\rm MRP}$\fi}
\def\ratem{\ifmmode {n_{\rm MRP}\over\tau_{\rm MRP}} \else
${n_{\rm MRP}\over\tau_{\rm MRP}}$\fi} 
\def\ratel{\ifmmode {n_{\rm LMXB}\over\tau_{\rm LMXB}} \else
${n_{\rm LMXB}\over\tau_{\rm LMXB}}$\fi} 
\def\LX{\ifmmode L_X \else $L_X$\fi} 
\def\LB{\ifmmode L_B \else $L_B$\fi} 

\begin{document}

\title{Cosmic Star Formation History and Deep X-ray Imaging in the 
\emph{XMM-Newton} and \emph{Chandra} Era}

\author{Pranab\,Ghosh\inst{}}

\institute{Department of Astronomy \& Astrophysics, Tata Institute of 
Fundamental Research, Bombay 400 005, INDIA}

\maketitle 

\begin{abstract}

I summarize X-ray diagnostic studies of cosmic star formation in terms of 
evolutionary schemes for X-ray binary evolution in normal galaxies with 
evolving star formation. Deep X-ray imaging studies by \emph{Chandra} and
\emph{XMM-Newton} are beginning to constrain both the X-ray luminosity 
evolution of galaxies and the $\log N$--$\log S$ diagnostics of the X-ray 
background: I discuss these in the above context, summarizing current 
understanding and future prospects.    

\end{abstract}

\section{Introduction}

This is a brief account of the current status and future potentials of the 
X-ray diagnostics of the history of cosmic star-formation rate (SFR). 
Global SFR has undergone strong cosmological evolution: it was $\sim$ 10 
times its present value at $z\approx 1$, had a peak value $\sim$ 10--100 
times the present one in the redshift range $z\sim$ 1.5--3.5, and declined 
again at high $z$ (Madau, Pozzetti \& Dickinson 1998, henceforth M98; Blain, 
Smail, Ivison \& Kneib 1999, henceforth B99a; Blain \etal~1999, henceforth 
B99b, and references therein). Details of the SFR at high redshifts are 
still somewhat uncertain, because much of the star formation at $2\lo z\lo 
5$ may be dust-obscured and so missed by optical surveys, but detected 
readily through the copious submillimeter emission from the dust heated by 
star formation. 

The X-ray emission of a normal galaxy (\ie, one without an active nucleus) 
is believed to be dominated by the integrated emission of the galaxy's 
X-ray binary population: this statement may be somewhat dependent 
on the X-ray energy band (see Sec.~\ref{pghosh-E1_sec:future}), 
but current understanding does suggest
that, in the canonical 2--10 keV X-ray band, the statement is valid for 
most normal galaxies. I summarize in this paper recent studies made in 
collaboration with N. White, A. Ptak, and R. Griffiths (White \& Ghosh 
1998, henceforth WG98; Ghosh \& White 2001, henceforth GW01; Ptak 
\etal~2001, henceforth Ptak01) on the basic imprints of an evolving SFR 
on the evolution of X-ray binary populations of galaxies, on the general 
consequences of these studies for deep X-ray imaging of galaxy fields by 
\emph{Chandra} and \emph{XMM-Newton}, and on the first, specific results 
that have emerged so far on the X-ray luminosity evolution in the Hubble
Deep Field (\object{HDF}), and on the $\log N$--$\log S$ diagnostics of 
the X-ray background. First results of Brandt \etal~(2001, henceforth 
Bran01) from the $\sim$ 0.5 Ms \emph{Chandra} exposure of HDF North 
(\object{HDF-N}) suggest an evolution of the X-ray luminosities, \LX , 
of bright spirals from the Local Universe to $z\approx 0.5$, which I
compare with the GW01 predictions from current SFR models: I also discuss 
the roles of global and individual SFR histories in this context. 
Fluctuation analyses of the $\sim$ 1 Ms \emph{Chandra} exposure of 
(\object{HDF-N}) suggest (Miyaji \& Griffiths 2002, henceforth MG02) that 
the $\log N$--$\log S$ plot in the soft X-ray band continues to rise at
low ($S\sim 10^{-16}$ -- $10^{-17}$ erg cm$^{-2}$ s$^{-1}$) fluxes, 
indicating that the X-ray background at these fluxes is possibly 
dominated by a new population of faint X-ray sources rather than the
canonical integrated AGN population (Gilli \etal~2001, henceforth GSH),
whose contribution shows a cosmological flattening at these fluxes, and
so lies much below the observational values suggested by MG02: in view of
the Ptak01 predictions which I discuss, it is plausible that this 
additional population is, in fact, that of normal galaxies showing the
signature of their SFR histories through X-ray emission.  
             
\section{SFR profiles and X-ray luminosity evolution}
\label{pghosh-E1_sec:xevol}

In the approach of WG98 and GW01, the total X-ray output of a normal galaxy 
is modeled as the sum of those of its high-mass X-ray binaries (HMXB) and 
low-mass X-ray binaries (LMXB), the evolution of each species ``$i$'' being 
described by a timescale $\tau_i$. The effects of the dependence of $\tau_i$ 
on the binary period and other parameters are studied by running the 
evolutionary scheme over ranges of likely values of $\tau_i$ given in the 
literature. The evolution of the HMXB population in response to an evolving 
star-formation rate SFR(t) is given by
\begin{equation}
{\partial\hmxb(t)\over\partial t} = \alpha_h {\rm SFR}(t)-{\hmxb(t)
\over\tauh},
\label{eq:evohmxb}
\end {equation}
where \hmxb~is the number density of HMXBs in the galaxy, and
\tauh~is the HMXB evolution timescale. $\alpha_h$ is the rate of 
formation of HMXBs per unit SFR, given approximately by $\alpha_h = 
{1\over 2}f_{\rm binary}f^h_{\rm prim}f^h_{\rm SN}$, where $f_{\rm 
binary}$ is the fraction of all stars in binaries, $f^h_{\rm prim}$ is 
that fraction of primordial binaries which has the correct range of 
stellar masses and orbital periods for producing HMXBs (van den Heuvel 
1992, henceforth vdH92), and $f^h_{\rm SN}\approx 1$ is 
that fraction of massive binaries which survives the first supernova. 
In these calculations, a representative value $\tauh\sim 5\times 10^6$ 
yr is adopted according to current evolutionary models. Note that
\tauh~includes both (a) the time taken ($\sim 4-6\times 10^6$ yr) by the 
massive companion  of the neutron star to evolve from the instant of the 
neutron-star-producing supernova to the instant when the ``standard'' 
HMXB phase begins, and, (b) the (much shorter) duration ($\sim 2.5\times 
10^4$ yr) of this HMXB phase (vdH92 and references therein).

Of the two mechanisms of LMXB production generally envisaged, 
\emph{viz.},(a) production in cores of globular clusters due to tidal 
capture, and, (b) general production by evolution of primordial binaries, 
I describe here only the latter one (which must be the dominant mechanism 
at least for spiral galaxies, since globular-cluster LMXB populations of 
such galaxies can account only for relatively small fractions of their 
total X-ray luminosities), deferring the former to 
Sec.~\ref{pghosh-E1_sec:future}.
LMXB evolution from primordial binaries has two stages (WG98) 
after the supernova produces a post-supernova binary (PSNB) containing
the neutron star. First, the PSNB evolves on a timescale \taup~due to 
nuclear evolution of the neutron star's low-mass companion and/or 
decay of binary orbit due to gravitational radiation and magnetic 
braking, until the companion comes into Roche lobe contact and the LMXB 
turns on. Subsequently, the LMXB evolves on a timescale \tauel. Since 
\taup~and \tauel~are comparable in general, the two stages are 
described separately (WG98) by:
\begin{equation}
{\partial\psnb(t)\over\partial t} = \alpha_l {\rm SFR}(t)
-{\psnb(t)\over\taup},
\label{eq:evopsnb}
\end{equation}
\begin{equation}
{\partial\lmxb(t)\over\partial t} = {\psnb(t)\over\taup}
-{\lmxb(t)\over\tauel},
\label{eq:evolmxb}
\end{equation}
Here, \psnb~and \lmxb~are the respective number densities of PSNB and
LMXB in the galaxy, and $\alpha_l$ is the rate of formation of LMXB per 
unit SFR, given approximately by $\alpha_l = {1\over 2}f_{\rm binary}
f^l_{\rm prim}f^l_{\rm SN}$, the individual factors having meanings 
closely analogous to those for HMXBs (see GW01).

Evolution is displayed in terms of the redshift $z$, which is related to
the cosmic time $t$ by $t_9 = 13(z+1)^{-3/2}$, where $t_9$ is $t$ in
units of $10^9$ yr, and a value of $H_0=50$ km s$^{-1}$ Mpc$^{-1}$ has
been used\footnote{For ease of comparison with WG98, M98, and GW01, I 
use here a Friedman cosmology with $q_0 = 1/2$. Other values of the 
Hubble constant lead to a straightforward scaling: for $H_0=70$ km 
s$^{-1}$ Mpc$^{-1}$, for example, $t_9 \approx 10(z+1)^{-3/2}$, so that 
the results remain unchanged if all timescales are shortened by a factor 
of 1.3.}. I consider the suite of current SFR models detailed in Table 
1 to cover a plausible range, using the parameterization of B99a,b. 
Models of the ``peak'' class have the form:

\begin{equation}
{\rm SFR}_{\rm peak}(z) = 2\left(1+\exp{z\over z_{max}}
\right)^{-1}(1+z)^{p+{1\over 2z_{max}}},
\label{eq:SFRpeak}
\end{equation}
while those of the ``anvil'' class have the form:
\begin{equation}
{\rm SFR}_{\rm anvil}(z) = \cases{(1+z)^p,&{$z\leq z_{max}$},\cr
                                  (1+z_{max})^p,&{$z> z_{max}$}.\cr}
\label{eq:SFRanvil}
\end{equation}

\begin{table}[ht]
\caption{Star Formation Rate (SFR) Profiles}
\label{pghosh-E1_tab:tab1}
\footnotesize
\halign{
\rm#\hfil&\qquad\rm#\hfil&\qquad\rm#\hfil
&\qquad\rm#\hfil&\qquad\hfil\rm#\qquad\hfil\cr
\noalign{\vskip6pt}\noalign{\hrule}\noalign{\vskip2pt}
\noalign{\hrule}\noalign{\vskip6pt}
Model&$z_{max}$&$p$&Comments\cr
\noalign{\vskip6pt}\noalign{\hrule}\noalign{\vskip6pt}
Peak-M&0.39&4.6&Madau profile\cr
Hierarchical&0.73&4.8&Hierarchical Clustering\cr
Anvil-10&1.49&3.8&Monolithic Models\cr
Peak-G&0.63&3.9&Peak part of composite\cr
&&&``Gaussian'' Model\cr
Gaussian&N/A&N/A&Gaussian starburst\cr
&&&added at high $z$\cr
\noalign{\vskip6pt}\noalign{\hrule}\noalign{\vskip12pt}}
\end{table}

These functional forms are convenient since they have a
convenient low-$z$ limit, SFR$(z)\propto (1+z)^p$, where all SFR profiles 
must agree with the  optical/UV data (M98), and since the model
parameters can be manipulated to mimic a wide range of star-formation
histories (B99b). Peak-class profiles are useful for describing (a) SFRs
determined from optical/UV observations, \ie, Madau-type (M98) profiles,
called ``Peak-M'' in Table 1, and, (b) more general SFRs with enhanced 
star formation at high $z$, an example of which is the 
``hierarchical'' model of B99b, wherein the submillimeter emission is 
associated with galaxy mergers in an hierarchical clustering model. 
Anvil-class profiles are useful for describing the 
results of ``monolithic'' models. The ``Gaussian'' model (B99a,b)is an 
attempt at giving a good account of the SFR at both low and high $z$ by 
making a composite of the Peak-G model (see Table 1) and a Gaussian
starburst at a high redshift $z_p$, \ie, a component
\begin {equation}
{\rm SFR}_{\rm Gauss}(z) = \Theta\exp\left\{-{[t(z)-t(z_p)]^2\over
2\sigma^2}\right\}.
\label{eq:SFRGauss}
\end{equation}
Based on the $IRAS$ luminosity function, this componenet is devised to 
account for the high-$z$ data, particularly the submillimeter 
observations (B99a). For its parameters (see Table 1), I have used the 
revised values given in B99b (see GW01). In all models described here, 
no galaxies exist for sufficiently large redshifts, $z>$ 10.
\begin{figure}[ht]
  \begin{center}
    \epsfig{file=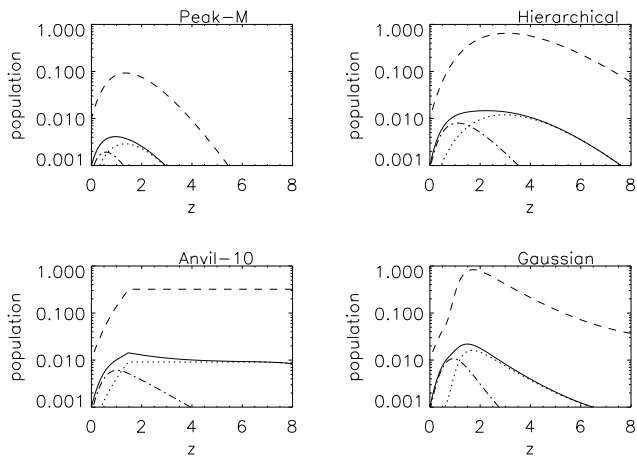, width=9cm}
\end{center}
\caption{Evolution of HMXB population (dotted line), LMXB population
(dash-dotted line), and the total X-ray luminosity \LX~(solid line) of a
galaxy with various SFR profiles (dashed line), from GW01. The effects of 
SFR variation are shown by keeping the evolutionary timescales fixed at 
\taup~= 1.9 Gyr and \tauel~= 1.0 Gyr for all cases, and choosing various
SFR profiles from Table 1. Each panel is labeled by the name of its SFR 
profile.} 
\label{pghosh-E1_fig:fig1}
\end{figure}
Figure 1 shows the prompt evolution of HMXBs and the slow evolution 
of LMXBs, and the evolution of the total X-ray binary population, where 
the two components have been so weighted as to represent the total X-ray 
emission from the galaxy. The HMXB profile closely follows the SFR profile 
because \tauh~is small compared to the SFR evolution timescale. By 
contrast, the LMXB profile has a significant lag behind the SFR profile 
because \taup~and \tauel~are comparable to SFR evolution timescale: the 
LMXB profile generally peaks at redshifts $\sim$ 1 -- 3 later than the HMXB 
profile --- a charcteristic signature of SFR evolution (WG98). Effects of 
both (a) varying the evolutionary timescales for fixed SFR profiles, and, 
(b) varying the SFR profile for fixed evolutionary timescales have been
studied: see GW01 for details. I display the latter variation in Figure 1 
to emphasize that, since, for sufficiently \emph{slow} LMXB evolution, the 
galaxy's X-ray emission is dominated by LMXBs at low redshifts ($0\lo z\lo 
1$), and by HMXBs at high redshifts, the total \LX -profile is 
strongly influenced at high redshifts by the SFR profile. Thus, 
determination of the \LX -profile even upto moderate redshifts may put 
interesting constraints on the SFR, making this an {\it independent\/} 
X-ray probe of cosmic star-formation history.

From their stacking analysis (see Bran01 and references therein for an
exposition of the technique), Bran01 estimate that the average X-ray 
luminosity of the bright spiral galaxies at an average redshift $z\approx 
0.5$ used in their study is about a factor of 3 higher than that in the 
local Universe. This observed evolution, \LX (0.5)/\LX (0.0) $\sim$ 3, can 
be compared with the theoretical results in Table 2. The degree of evolution 
from $z=0$ to $z=$ 0.5--1.0 increases from Madau-type profiles to those 
with additional star formation at high redshifts, the numbers for the 
Peak-M profile being in best agreement with Bran01.  
\begin{table}[ht]
\caption{Evolution of X-ray Luminosity \LX}
\label{pghosh-E1_tab:tab2}
\footnotesize
\halign{%
\rm#\hfil&\qquad\rm#\hfil&\qquad\rm#\hfil
&\qquad\rm#\hfil&\qquad\hfil\rm#\qquad\hfil\cr
\noalign{\vskip6pt}\noalign{\hrule}\noalign{\vskip2pt}
\noalign{\hrule}\noalign{\vskip6pt}
Model&$\taup$&$\tauel$&${\LX(0.5)\over\LX(0.0)}$&
${\LX(1.0)\over\LX(0.0)}$\cr
\noalign{\vskip6pt}\noalign{\hrule}\noalign{\vskip6pt}
Peak-M&1.9&0.1&3.9&5.4\cr
Peak-M&0.9&0.5&4.6&6.8\cr
Peak-M&1.9&1.0&3.4&4.1\cr
Hierarchical&1.9&1.0&6.2&11.3\cr
Anvil-10&1.9&1.0&5.4&10.1\cr
Gaussian&1.9&1.0&7.5&16.0\cr
\noalign{\vskip6pt}\noalign{\hrule}\noalign{\vskip12pt}}
\end{table}

\section{Global and individual SFRs}
\label{pghosh-E1_sec:global}

A new development in SFR research in the last 
three years has been the study of star-formation histories of individual 
galaxies and various galaxy-types. SFR profiles of individual galaxies, 
ranging from those in the Local Group to those in the \object{HDF} at 
redshifts $0.4\lo z\lo 1$, have been inferred, using a variety of 
techniques. For various galaxy-types, models of spectrophotometric 
evolution, which use the synthesis code \emph{P\'{e}gase} and are 
constrained by deep galaxy counts, have been developed (Rocca-Volmerange 
and Fioc 2000,henceforth RF00), leading to a model SFR profile for each 
type. In the light of these developments, let me now suggest what may be 
the true significance of the Bran01 results discussed above.
 
Bran01 used bright spirals for their stacking analysis. RF00 have shown 
that the model SFR profile for such (Sa-Sbc) spirals 
rises roughly in a Madau fashion from $z=0$ to $z\approx 1$ (which 
these authors ascribe to a bias in the original sample used to construct 
the Madau profile towards bright spirals), and thereafter flattens to a 
roughly constant value $\sim$ 12 times that at $z=0$, falling again at 
$z\go 7$. In the range $0<z\lo 7$, this profile can be roughly represented 
by an anvil-type profile (see Sec.~\ref{pghosh-E1_sec:xevol}),
with the parameter $z_{max}$ as given in Table 1, and the parameter
$p\approx$ 2.7. For such a profile with the timescales \taup~= 1.9 Gyr, 
\tauel~= 1.0 Gyr, as in Figure 1, the GW01 evolutionary scheme gives  
\LX (0.5)/\LX (0.0) = 3.3, and \LX (1.0)/\LX (0.0) = 5.4, in good 
agreement with both the Bran01 results and the Peak-M results given in
Table 2. It is now easy to see to see why the Peak-M profile would appear
to give a good account of the Bran01 results. In effect, the Bran01 
analysis may be probing the SFR profile of \emph{only} the bright spirals 
in HDF-N, and the fact that the Peak-M profile is consistent with the 
Bran01 results does \emph{not} imply that the global SFR necessarily 
follows the Peak-M profile.

\section{$\log N$--$\log S$ diagnostics: X-ray background}
\label{pghosh-E1_sec:lognlogs}

Based on the results of Sec.~\ref{pghosh-E1_sec:xevol}, Ptak01 calculated
the X-ray flux distributions and source count ($\log N$--$\log S$) plots 
expected for \object{HDF-N}. Figure 2 shows the Ptak01 plot in the soft 
(0.5--2.0 keV) X-ray band, which has proved to be a valuable diagnostic 
of current population synthesis models of the X-ray background, as I 
now summarize. 
\begin{figure}[ht]
  \begin{center}
    \epsfig{file=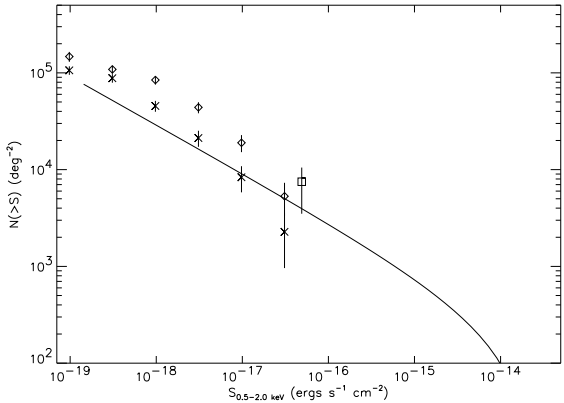, width=8cm}
\end{center}
\caption{$\log N$--$\log S$ plot in the soft (0.5 -- 2.0 keV) band for 
\object{HDF-N}, from Ptak01. The diamonds correspond to the Gaussian
SFR profile described in Sec.~\ref{pghosh-E1_sec:xevol}, and the crosses
to the Peak-M profile. Note that an interpolation through the former 
points is represented by a dashed line in Fig.2 of MG02 and Fig.2 
(bottom) of H02, and that through the latter points by a dotted line. 
The solid line here is the double power law fit of Tozzi \etal~to the
\emph{Chandra} observations of HDF South.} 
\label{pghosh-E1_fig:fig2}
\end{figure}
Hasinger (2002, henceforth H02) reminds us that the cosmic X-ray 
background has, in this \emph{Chandra} and \emph{XMM-Newton} era of deep 
X-ray surveys, been largely resolved into contributions from individual 
sources, the resolved fraction being $\go$ 90\% in the soft 
(0.5--2.0 keV) band, and similar in the harder (2--10 keV) band. 
The long-standing belief that these sources 
are predominantly active galactic nuclei (AGN), both unobscured and 
obscured (the so-called QSO-2s), was supported by the (now completed) 
optical identification programme which followed up the ROSAT deep survey, 
since it found the counterparts to be predominantly AGN. Ongoing optical 
identifications of the deepest \emph{Chandra} and \emph{XMM-Newton} fields 
are still far from complete. AGN population-synthesis models of the X-ray 
bacground are currently very useful and popular: these have been developed 
to a degree of sophistication and detail (see 
GSH, which has references to earlier models) sufficient for extracting 
information about AGN population properties. The recent, ultradeep ($\sim$ 
1 Ms) observations of both \object{HDF-N} and the Chandra Deep Field South 
(\object{CDFS}) have led to $\log N$--$\log S$ plots in the soft (0.5--2.0 
keV) X-ray band which go down to fluxes $S\sim 5\times 10^{-17}$ erg 
cm$^{-2}$ s$^{-1}$: these are fitted well by the GSH models, which show a 
clear cosmological flattening at fluxes below the above limit (H02; MG02).  

Fluctuation analysis is a powerful tool for constraining the source counts
below source detection limit (see MG02 and references therein for an 
exposition of the method), which has been successfully tested on 
data from previous X-ray missions. Its recent application by MG02 to the 1 
Ms observation of \object{HDF-N} has yielded the remarkable result that 
the constraints so obtained on the soft-band $\log N$--$\log S$ plot  
suggest that the extension of the plot down to fluxes $S\sim 7\times 
10^{-18}$ erg cm$^{-2}$ s$^{-1}$ continues to rise as at higher fluxes,
showing no signs of the cosmological flattening characteristic of the GSH
models (MG02; H02). The most obvious interpretaion is that, while the AGN 
contribution, as modelled by GSH, begins to saturate at these fluxes, a 
new population of faint sources begins to dominate. The fact that the 
the extension of the $\log N$--$\log S$ plot, as inferred from the 
fluctuation-analysis constraints of MG02, agree well with that shown in 
the above Ptak01 plot (since the figure showing this appears twice in 
these proceedings, Fig.2 in MG02 and Fig.2 (bottom) in H02, I do not repeat 
it here), particularly for the Gaussian SFR profile, therfore opens 
the exciting possibility that first signatures of cosmic star formation 
in the soft X-ray band $\log N$--$\log S$ plots are revealing themselves.      

\section{Future prospects}
\label{pghosh-E1_sec:future}          

Even in this \emph{Chandra} and \emph{XMM-Newton} era, truly new results 
on \LX -evolution and SFR signature have been possible so far only by 
going below the source detection limit with special techniques like 
stacking and fluctuation analysis. These are suggestive indications, which
must be confirmed with source detection at lower fluxes, first with longer
exposures with \emph{Chandra} and \emph{XMM-Newton}, and then with the 
next generation of satellites like \emph{Constellation}\emph{-X} and 
\emph{XEUS}.
On the theoretical side, the evolutionary scheme must be generalized to
include several additional effects, \eg, (a) in the \emph{soft} X-ray band, 
the output of a normal galaxy may have very significant contributions from
supernova remnants (G. Hasinger, personal communication), and, (b) tidal 
capture creation of LMXBs in globular clusters may be the dominant
production mechanism in certain galaxy-types. Inclusion of these effects 
presents no difficulties of principle, and is now under way: the results 
will be described elsewhere.


\begin{thebibliography}{}

\bibitem[\protect\astroncite{}{}]{pghosh-E1:B99a}
Blain, A. W., Smail, I., Ivison, R. J. \& Kneib, J.-P. 1999,
MNRAS 302, 632 (B99a)
\bibitem[\protect\astroncite{}{}]{pghosh-E1:B99b}
Blain, A. W., Jameson, A., Smail, I., Longair, M. S., Kneib, J.-P.
\&  Ivison, R. J. 1999, MNRAS 309, 715 (B99b)
\bibitem[\protect\astroncite{}{}]{pghosh-E1:Bran01}
Brandt, W. N. \etal, 2001, AJ 122, 1 [astro-ph/0102411] (Bran01)
\bibitem[\protect\astroncite{}{}]{pghosh-E1:GW01}
Ghosh, P. \& White, N. 2001, ApJ 559, L97 (GW01)
\bibitem[\protect\astroncite{}{}]{pghosh-E1:GSH}
Gilli, R., Salvati, M., \& Hasinger, G. 2001, A\&A 366, 407 (GSH)
\bibitem[\protect\astroncite{}{}]{pghosh-E1:H02}
Hasinger, G. 2002, these proceedings [astro-ph/0202430] (H02)
\bibitem[\protect\astroncite{}{}]{pghosh-E1:M98}
Madau, P., Pozzetti, L., \& Dickinson, M. 1998, MNRAS 498, 106 (M98)
\bibitem[\protect\astroncite{}{}]{pghosh-E1:MG02}
Miyaji, T. \& Griffiths, R. 2002, these proceedings 
[astro-ph/0202048] (MG02)
\bibitem[\protect\astroncite{}{}]{pghosh-E1:Ptak01}
Ptak, A., Griffiths, R., White, N. \& Ghosh, P. 2001, ApJ 559, L91 
(Ptak01)
\bibitem[\protect\astroncite{}{}]{pghosh-E1:RF00}
Rocca-Volmerange, B. \& Fioc, M. 2000, in Toward A New Millenium in
Galaxy Morphology, ed. D. L. Block, I. Puerari, A. Stockton \& D.
Ferreira, Kluwer: Dordrecht [astro-ph/0001398] (RF00)
\bibitem[\protect\astroncite{}{}]{pghosh-E1:WG98}
White, N. \& Ghosh, P. 1998, ApJ 504, L31 (WG98)
\bibitem[\protect\astroncite{}{}]{pghosh-E1:vdH92}
van den Heuvel, E. P. J. 1992, in X-ray Binaries and Recycled 
Pulsars, ed. E. P. J. van den Heuvel and S. A. Rappaport, Kluwer: 
Dordrecht, p.233 (vdH92)

\end{thebibliography}
\end{document}